 \documentclass[journal,10pt]{IEEEtran}

\usepackage{mathtools}
\usepackage{amsmath} 
\usepackage{amssymb}
\usepackage{amsfonts}
\usepackage{verbatim} 
\usepackage{bm,color,soul}
\usepackage{algorithm,algorithmic}
\usepackage{cite}
\usepackage{caption}
\usepackage{subfigure}
\usepackage{stfloats} 
\usepackage[colorlinks, linkcolor=black, anchorcolor=black, citecolor=black]{hyperref} 
\usepackage{subeqnarray}
\usepackage{cases}
\usepackage{amsmath}
\usepackage{makecell} 
\usepackage{enumerate}

\newcommand{\rnum}[1]{\uppercase\expandafter{\romannumeral #1\relax}}
\ifCLASSINFOpdf
\else
\fi
\makeatletter
\renewcommand{\maketag@@@}[1]{\hbox{\m@th\normalsize\normalfont#1}}
\usepackage{array}
\usepackage{url}
\usepackage{balance}
\IEEEoverridecommandlockouts

\begin{document}
%
\title{Energy-Efficient Transmit Beamforming and Antenna Selection with Non-Linear PA Efficiency}

\author{\IEEEauthorblockN{Yuan Fang, Yi Huang, Chuan Ma, Yinghao Jin, Gaoyuan Cheng, Guanlin Wu, and Jie Xu\vspace{-0.5cm}
        }

\thanks{Y. Fang, G. Cheng, G. Wu, and J. Xu are with the School of Science and Engineering (SSE) and the Future Network of Intelligence Institute (FNii), The Chinese University of Hong Kong, Shenzhen, Shenzhen 518172, China (e-mail: fangyuan@cuhk.edu.cn; gaoyuancheng@link.cuhk.edu.cn; guanlinwu1@link.cuhk.edu.cn; xujie@cuhk.edu.cn). J. Xu is the corresponding author.} 
\thanks{Y. Huang is with the Department of Information and Communication Engineering, Tongji University, China (e-mail: huangyi718b@tongji.edu.cn).}
\thanks{C. Ma and Y. Jin are with Huawei Technologies Co., Ltd., Shanghai 200000, China (e-mail: machuan2@huawei.com; jinyinghao@huawei.com). }
}

\maketitle
\begin{abstract}
This letter studies the energy-efficient design in a downlink multi-antenna multi-user system consisting of a multi-antenna base station (BS) and multiple single-antenna users, by considering the practical non-linear power amplifier (PA) efficiency and the on-off power consumption of radio frequency (RF) chain at each transmit antenna. Under this setup, we jointly optimize the transmit beamforming and antenna on/off selection at the BS to minimize its total power consumption while ensuring the individual signal-to-interference-plus-noise ratio (SINR) constraints at the users. However, due to the non-linear PA efficiency and the on-off RF chain power consumption, the formulated SINR-constrained power minimization problem is highly non-convex and difficult to solve. To tackle this issue, we propose an efficient algorithm to obtain a high-quality solution based on the technique of sequential convex approximation (SCA). We provide numerical results to validate the performance of our proposed design. It is shown that at the optimized solution, the BS tends to activate fewer antennas and use higher power transmission at each antenna to exploit the non-linear PA efficiency. 
\end{abstract}

\begin{IEEEkeywords}
 Energy-efficient communication, non-linear PA efficiency, transmit beamforming, antenna selection.
\end{IEEEkeywords}

%
\IEEEpeerreviewmaketitle

\section{Introduction}
Fifth-generation (5G) and beyond cellular networks are expected to provide ultra-high data-rate throughput, ultra-high reliability, and ultra-low latency, by exploiting various new technologies such as extremely large-scale multiple-input and multiple-output (MIMO) \cite{cui2022channel}. This, however, may lead to tremendous energy consumption and operational cost, as well as significant carbon emissions \cite{mao2021ai}. Extensive research efforts have been conducted to achieve green and carbon-neutral communications from different perspectives such as energy-efficient transmit beamforming and antenna selection \cite{mehanna2013joint,jiang2012antenna,luo2014downlink}, user association and base station (BS) activation \cite{liao2014base,shi2014group}, as well as renewable-powered BSs \cite{xu2015cost}. 

Among others, the joint transmit beamforming and antenna on/off selection has attracted a lot of research interest to enhance the energy efficiency (EE) of cellular networks with multi-antenna BSs, in which the BS properly designs the transmit beamforming and switches off some transmit antennas to save the  power consumption at the associated radio-frequency (RF) chains. For instance, the work \cite{jiang2012antenna} considered a point-to-point MIMO system with one single data stream, in which the transmit power control and antenna selection are jointly optimized to maximize the bits-per-Joule EE. Furthermore, the authors in \cite{he2016joint} and \cite{xu2013energy} studied downlink multiuser multi-antenna systems by considering the linear transmit beamforming and capacity-achieving dirty paper coding, respectively, in which the transmit beamforming/precoding and antenna selection are jointly designed to maximize the EE. In addition, the works \cite{li2014energy} and \cite{li2014energy} considered massive MIMO systems, in which antenna selection algorithms are developed to enhance the EE based on the ideas of binary search and particle swarm, respectively. Moreover, such designs have been extended to enhance the EE of extremely large-scale MIMO systems in \cite{marinello2020antenna}. Despite the research progresses, these prior works normally assumed fixed power amplifier (PA) efficiency at transmit antennas. This, however, may not be valid in practice and make the corresponding designs less energy efficient.

In practice, the PAs at a BS account for about 50\%-80\% of its total power consumption, and the efficiency of each PA is highly non-linear in general \cite{bogucka2011degrees}. In particular, the maximum PA efficiency is normally achieved when the PA input signal power reaches a saturation point at which the maximum output signal power is achieved. When the PA input signal power deviates from such saturation point, the PA efficiency may drop significantly, thus leading to rapidly decreased system EE \cite{cripps2006rf,joung2013spectral}. As a result, the non-linear PA efficiency may seriously affect the EE of multi-antenna systems, and the conventional designs assuming linear PA efficiency may not be energy efficient in general. This thus motivates our investigation in this work. 

In this letter, we investigate the energy-efficient joint transmit beamforming and antenna (on/off) selection in a downlink multi-antenna communication system with one multi-antenna BS and multiple single-antenna users, by considering both the non-linear PA efficiency and the on-off power consumption of RF chain at each antenna. First, we introduce a general non-linear PA efficiency model and the corresponding power consumption model for the BS. Next, we formulate the joint transmit beamforming and antenna on/off selection design problem, with the objective of minimizing total power consumption at the BS subject to the minimum signal-to-interference-plus-noise ratio (SINR) at the users and the per-antenna transmit power constraints at the BS. The formulated problem is highly non-convex due to the non-linear PA efficiency and the binary on-off RF chain power consumption, and thus is difficult to solve. To overcome this issue, we propose an efficient method by utilizing  the technique of sequential convex approximation (SCA), together with a beamforming weight-based antenna selection algorithm. Finally, we provide numerical results to validate the performance of our proposed design. It is shown that the BS tends to activate fewer antennas but use higher transmit power at each antenna, which is different from the conventional design with a fixed PA efficiency. It is also shown that the proposed method significantly reduces the power consumption at the BS as compared to the conventional designs considering the fixed PA efficiency and/or ignoring the on-off RF chain power consumption.

\section{System Model}
\subsection{Signal Model}
\begin{figure}[htbp]
	\setlength{\abovecaptionskip}{-0pt}
	\setlength{\belowcaptionskip}{-8pt}
	\centering
	\includegraphics[width= 0.42\textwidth]{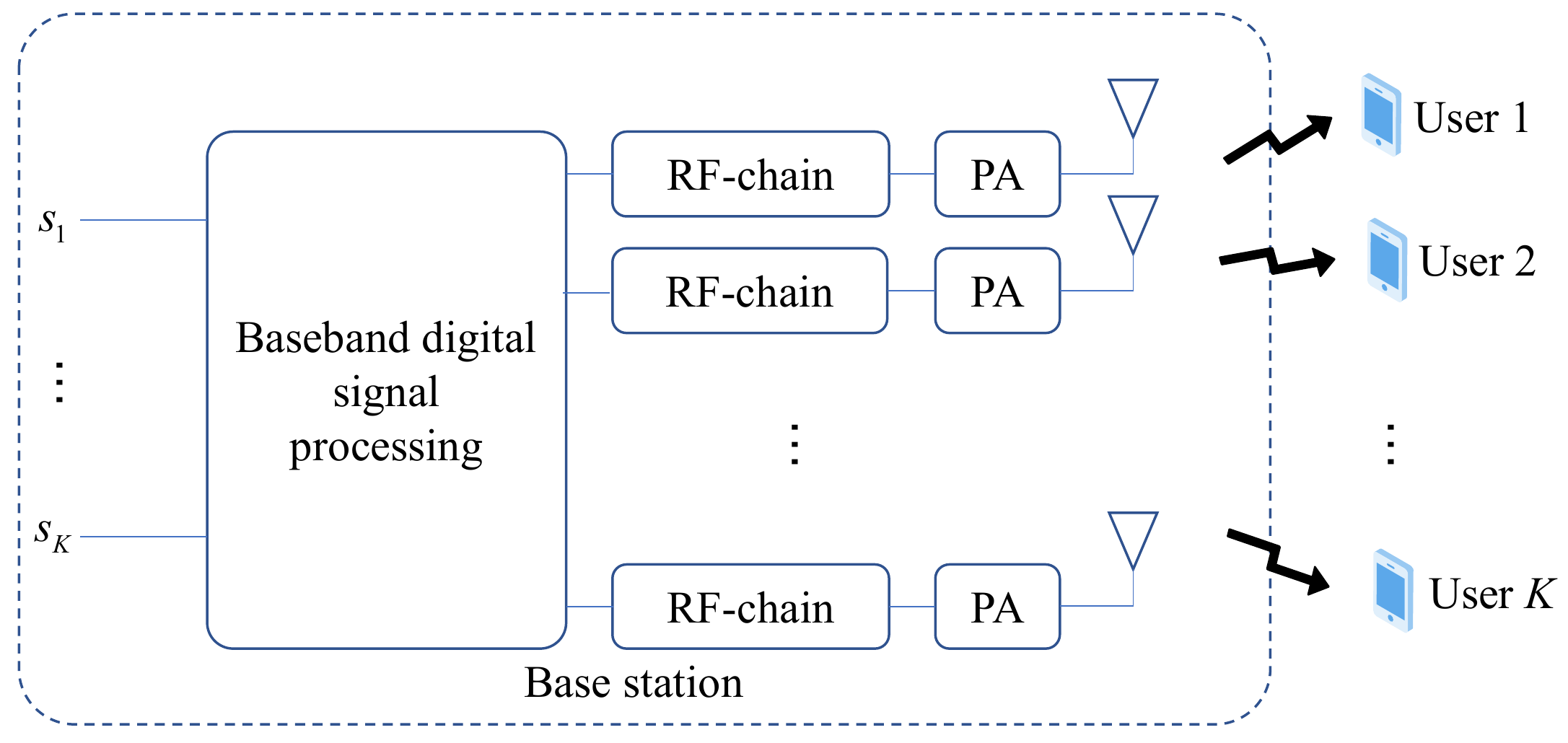}
	\DeclareGraphicsExtensions.
	\caption{The system configuration.}
	\label{SystemModel}
\end{figure}
We consider a downlink multi-antenna multiuser system consisting of one BS with $N$ antennas and $K$ users each with one single antenna, as shown in Fig. \ref{SystemModel}. Let $\mathcal{N} = \{ 1, \ldots ,N\} $ and $\mathcal{K} = \{ 1, \ldots ,K\} $ denote the set of transmit antennas at the BS and that of users, respectively. Let $s_k$ denote the desired signal for user $k$, which is a random variable with zero mean and unit variance, i.e., $\mathbb{E}[ {{{\left| {s_k} \right|}^2}} ] = 1$, with $\mathbb{E}[\cdot]$ denoting the statistic expectation. By using the transmit beamforming, the baseband signal transmitted from the BS is expressed as
\begin{align}\label{transmit_signal}
        {\boldsymbol{x}} = \sum\limits_{k \in \mathcal{K}} {{\boldsymbol{w}}_{k}s_k}, 
\end{align}
where ${\boldsymbol{w}}_{k} \in {\mathbb{C}^{{N} \times 1}}$ denotes the beamforming vector for user $k$. Let ${\boldsymbol{h}}_{k}^{H} \in {\mathbb{C}^{1 \times {N}}}$ denote the channel vector from the BS to user $k$. Then, the received signal at user $k$ is given by
\begin{align}
        y_k = {\boldsymbol{h}_{k}^{H}{\boldsymbol{w}}_{k}s_k} + \sum\limits_{i \ne k} {{\boldsymbol{h}_{k}^{H}{\boldsymbol{w}}_{i}s_i} }  + z_k,
\end{align}
where $z_k$ denotes the additive white Gaussian noise (AWGN) and at the receiver of each user $k$ that is a circularly symmetric complex Gaussian (CSCG) random variable with mean zero and variance $\sigma^2$. Hence, the SINR at user $k$ is given by
\begin{align}\label{SINR}
        \gamma_k = \frac{{{{\left| {\boldsymbol{h}_{k}^{H}{\boldsymbol{w}}_{k}}  \right|}^2}}}{{\sum\limits_{i \ne k} {{{\left| {\boldsymbol{h}_{k}^{H}{\boldsymbol{w}}_{i}}  \right|}^2}}  + \sigma _k^2}}.
\end{align}
\subsection{Power Consumption Model}
Next, we consider the power consumption at the BS, which consists of the non-linear power consumption by the PAs, the on-off power consumption by RF chains, and static power consumption by other components, which are detailed in the following, respectively.
\subsubsection{Non-linear  Power Consumption by PAs}
In practice, the efficiency of PA at each antenna $n \in \mathcal N$ is non-linear, which can be expressed as \cite{cripps2006rf}
\begin{align}
        {\eta _{n}}{=}{\eta^{\text{max}}}{\left( {\frac{{P_{n}^{{\text{out }}}}}{{P_{n}^{\max }}}} \right)^\beta },0 < \beta  \leq 1,
\end{align}
where $P_{n}^{\max }$ is maximum transmit power or output power of PA at antenna $n$, $P_{n}^{{\text{out }}}$ is the transmit signal power on antenna $n$, ${\eta^{\text{max}}}$ is the maximum efficiency of the PA, and $\beta$ is the efficiency factor depending on the specific type of PA. For instance, we have  $\beta = 1$ for class-A PA and $\beta = 0.5$ for class-B PA \cite{cripps2006rf}. Notice that based on the signal model in \eqref{transmit_signal}, we have ${P_{n}^{{\text{out }}}} = \sum\nolimits_{k \in \mathcal{K}} {\left| {{\boldsymbol{1}_n}\boldsymbol{w}_{k}} \right|}^2$, where $\boldsymbol{1}_n\in {\mathbb{C}^{{1} \times  {N}}}$ is a vector with its $n$-th element being one and others zero. Hence, the power consumption of the $N$ PAs at the BS is given by
\begin{align} 
        P_{\text{PA}}(\{\boldsymbol{w}_{k}\})&=\sum\limits_{n \in \mathcal{N}} {\frac{1}{{{\eta _{n}}}}P_{n}^{{\text{out }}}} = \sum\limits_{n \in \mathcal{N}} \frac{1}{{{\eta^{\text{max}}}}} {\left( {P_{n}^{\max }} \right)}^\beta \left(P_{n}^{{\text{out }}}\right)^{1-\beta} \nonumber\\
        &= {\frac{1}{{{\eta^{\text{max}}}}}\sum\limits_{n \in \mathcal{N}} {{{\left( {P_{n}^{\max }} \right)}^\beta }{{\left( {\sum\limits_{k \in \mathcal{K}} { {{{\left| {{\boldsymbol{1}_n}\boldsymbol{w}_{k}} \right|}^2}} } } \right)}^{1 - \beta }}} }.\label{PA_power_consum}
\end{align}
Furthermore, suppose that the maximum transmit power at each antenna n is given by $P_n^{\text{max}}$, and the maximum sum transmit power is given by $P_t$. Then we have 
\begin{align}
        &P_{n}^{{\text{out }}} = \sum_{k\in \mathcal K} {\left| {{\boldsymbol{1}_n}\boldsymbol{w}_{k}} \right|}^2 \leq P_{n}^{\max },\label{per-antenna_power}\\
        &\sum\limits_{k \in \mathcal{K}} \left\| {\boldsymbol{w}}_{k}\right\|^2 \leq P_{t}.\label{sum_tansmit_power}
\end{align}

\subsubsection{On-off Power Consumption by RF Chains}
Besides the non-linear power consumption by PAs, the power consumption by RF chains at the BS depends on the on-off status of each RF chain. In particular, if each antenna $n$ is on, i.e.,  antenna $n$ is transmitting with $P_{n}^{{\text{out }}} >0$, then the  associated RF chain needs to consume a fixed power consumption, given by 
\begin{align}
        {P_\text{RF}^{s}} = {P_\text{DAC}} + {P_\text{mix}} + {P_\text{filt}} + {P_\text{syn}},   
\end{align}
where ${P_\text{DAC}}$, ${P_\text{mix}}$, ${P_\text{filt}}$, and ${P_\text{syn}}$ represent power consumption by the digital-to-analog converter (DAC), mixer, filter, and synthesizer, respectively, at the RF chain. Otherwise, if antenna n is off. When the RF chain is on, i.e., the antenna is transmitting with $P_{n}^{{\text{out }}} >0$, the components on the RF chain are on with a fixed power consumption ${P_\text{RF}^{s}}$. Otherwise, with $P_{n}^{{\text{out }}} = 0$, then these components can be switched off, such that the power consumption becomes zero. By combining the $N$ RF-chains, the total RF chain power consumption is given by
\begin{align}
        P_{\text{RF}}(\{\boldsymbol{w}_{k}\})&= {{P_{\text{RF}}^{s}}\sum\limits_{n \in \mathcal{N}} {{\text{I}{\left\{ {\sum\limits_{k \in \mathcal{K}} {{{{\left| {{\boldsymbol{1}_n}\boldsymbol{w}_{k}} \right|}^2}} } } \right\}}}} },\label{RF_power_consum} 
\end{align}
where ${\text{I}}\left\{ \cdot \right\}$ denotes the indicator function which is defined by
\begin{align}
        {\text{I}}\left\{ x \right\} = \left\{ \begin{gathered}
                \begin{array}{*{20}{c}}
                0,& \text{if}\quad {x = 0} 
              \end{array} \hfill \\
                \begin{array}{*{20}{c}}
                1,& \text{otherwise} 
              \end{array} \hfill \\ 
              \end{gathered}  \right.
\end{align}
\subsubsection{Static Power Consumption}
In addition to the PAs and RF chains, other components at the BS such as backhauls, cooling systems, baseband processing, and power supply also consume energy. Their power consumption is generally static, and thus is modeled by a constant term $P_c$.

By combining the power consumption by PAs in \eqref{PA_power_consum}, RF chains \eqref{RF_power_consum}, and other components $P_c$, the total power consumption at the BS is given by 
\begin{align}\label{total_power_consum}
        P_{\text{BS}}^{\text{tot}} =  P_{\text{PA}}(\{\boldsymbol{w}_{k}\}) + P_{\text{RF}}(\{\boldsymbol{w}_{k}\}) + P_c.
\end{align}
\subsection{Problem Formulation}
Our objective is to minimize total power consumption at the BS in \eqref{total_power_consum}, while ensuring the minimum SINR requirement $\Gamma_k$ for each user $k\in \mathcal K$, the per-antenna transmit power constraints in \eqref{per-antenna_power}, and the sum transmit power constraint in \eqref{sum_tansmit_power}. The SINR-constrained total power minimization problem is formulated as
\begin{subequations}
\begin{align}
        &(\text{P1}):&{\mathop {\text{min}}\limits_{\left\{ {\boldsymbol{w}_{k}} \right\}} } \quad &  P_{\text{PA}}(\{\boldsymbol{w}_{k}\}) + P_{\text{RF}}(\{\boldsymbol{w}_{k}\}) +P_c \nonumber \\
        & &\text{s.t.}\quad &{ { {\frac{{{{\left| {{\boldsymbol{h}_{k}^{H}{\boldsymbol{w}}_{k}} } \right|}^2}}}{{\sum\limits_{i \ne k} {{{\left| {\boldsymbol{h}_{k}^{H}{\boldsymbol{w}}_{i}}  \right|}^2}}  + \sigma _k^2}}} \geq {\Gamma_k},\forall k \in \mathcal{K}} } \label{P1_cons1}\\
        &&&{\sum\limits_{k \in \mathcal{K}} { {{{\left| {\boldsymbol{1}_n\boldsymbol{w}_{k}} \right|}^2}} }  \leq {P_{n}^{\max }} , \forall n \in \mathcal{N}}\label{P1_cons2}\\
        &&&\sum\limits_{k \in \mathcal{K}} \left\| {\boldsymbol{w}}_{k}\right\|^2 \leq P_{t}.\label{P1_cons3}
\end{align}
\end{subequations}
 Note that in the objective function, the PA power consumption term is a concave function due to the non-linear PA efficiency, and the RF chain power consumption term is binary and thus non-convex due to the antenna on-off selection. Hence, problem (\text{P1}) is a non-convex problem and difficult to solve. 

Before solving problem (P1), we need to check its feasibility by solving the following feasibility problem: 
\begin{subequations}
        \begin{align}
                &(\text{P2}):&\text{Find} \quad & \{\boldsymbol{w}_{1},\ldots,\boldsymbol{w}_{K}\} \nonumber \\
                & &\text{s.t.}\quad &\eqref{P1_cons1}, \eqref{P1_cons2}, \eqref{P1_cons3}. \nonumber 
        \end{align}
        \end{subequations} 
 Notice that constraint \eqref{P1_cons1} is equivalent to the following convex second-order (SOC) constraint:
\begin{align}
        \sqrt{\sum\limits_{i \ne k}  \left|{\boldsymbol{h}_{k}^{H}{\boldsymbol{w}}_{i}}  \right|^2  + \sigma _k^2}  \leq \frac{1}{\sqrt{{\Gamma_k}}} \mathcal{R}\left({\boldsymbol{h}_{k}^{H}{\boldsymbol{w}}_{k}}  \right),\label{sinr_cons_eq}
\end{align}
where $\mathcal{R}(\cdot)$ denotes the real component of a complex number. By replacing \eqref{P1_cons1} as \eqref{sinr_cons_eq}, problem (P2) becomes a convex problem that can be optimally solved by standard convex solvers such as CVX \cite{grant2014cvx}. In the sequel, we focus on the case when problem (P1) is feasible.

\section{SCA-Based Solution to Problem (P1)}
In this section, we propose an efficient algorithm to solve problem (P1) by SCA together with a heuristic antenna selection design. 

First, we deal with the indicator function involved in the on-off RF chain power consumption. Towards this end, notice that the indicator function can be equivalently rewrite as \cite{sriperumbudur2011majorization}
\begin{align} \label{log:apx:thm}
        \text{I}\{x\}=\lim _{\epsilon \rightarrow 0} \frac{\log \left(1+x \epsilon^{-1}\right)}{\log \left(1+\epsilon^{-1}\right)}, x \geq 0.
\end{align} 
Accordingly, the indicator function can be approximated as $\frac{\log \left(1+x \epsilon^{-1}\right)}{\log \left(1+\epsilon^{-1}\right)}$ under a properly chosen $\epsilon$, where the approximation becomes more accurate when $\epsilon$ becomes smaller. As a result, the total RF chain power consumption is approximated as
\begin{align}
        P_{\text{RF}}(\{\boldsymbol{w}_{k}\}) \!\approx\! \hat P_{\text{RF}}(\{\boldsymbol{w}_{k}\}) \!=\!  {{P_{\text{RF}}^{s}}\!\!\sum\limits_{n \in \mathcal{N}}\!\! \frac{ {\log \left( {1 \!\!+\!\!  \epsilon^{ - 1} \!\!{\sum\limits_{k \in \mathcal{K}} \!\!{ {{{\left| {\boldsymbol{1}_n^{{N}}\boldsymbol{w}_{k}} \right|}^2}} } } } \right)}}{{\log \left( {1 + \epsilon^{ - 1}} \right)}} },\label{RF_power_consum_re} 
\end{align}
which is concave with respect to $\sum\nolimits_{k \in \mathcal{K}} {{{\left| {\boldsymbol{1}_n^{{N}}\boldsymbol{w}_{k}} \right|}^2}}$. As a result, problem (P1) is approximated as follows by omitting the constant term $P_c$ in the objective function:
\begin{subequations}
        \begin{align}
                &(\text{P3}):&{\mathop {\text{min}}\limits_{\left\{ {\boldsymbol{w}_{k}} \right\}} } \quad &  P_{\text{PA}}(\{\boldsymbol{w}_{k}\}) + \hat{P}_{\text{RF}}(\{\boldsymbol{w}_{k}\})\nonumber \\
                & &\text{s.t.}\quad &\eqref{P1_cons1}, \eqref{P1_cons2}, \eqref{P1_cons3}. \nonumber
        \end{align}
        \end{subequations}

Then, we use the SCA technique to solve (P3), in which (P3) is approximated into a series of convex problems in an iterative manner. Specially, we consider a particular iteration $j \ge 1$. Let $\{\boldsymbol{\bar w}_{k}^{\left( j \right)}\}$ be the local point of $\{\boldsymbol{w}_{k}\}$ at the $j$-th iteration of SCA. Note that any concave
function is globally upper-bounded by its first-order Taylor expansion at. Thus, with given local point $\{\boldsymbol{\bar w}_{k}^{\left( j \right)}\}$, the PA power consumption term ${P}_{\text{PA}}(\{\boldsymbol{w}_{k}\})$ in \eqref{PA_power_consum} is approximated by its upper bound $\tilde{P}_{\text{PA}}^{(j)}(\{\boldsymbol{w}_{k}\})$, i.e.,
 \begin{small}
\begin{align}\label{TransmitPower_sca_appr}
        \tilde{P}_{\text{PA}}^{(j)}(\{\boldsymbol{w}_{k}\})= &\sum\limits_{n \in \mathcal{N}} \frac{{{\left(  {P_{n}^{\max }} \right)}^\beta }}{{{\eta^{\text{max}}}}} {\left( {\sum\limits_{k \in \mathcal{K}} { {{{\left| {\boldsymbol{1}_n^{{N}}\boldsymbol{\bar w}_{k}^{\left( j \right)}} \right|}^2}} } } \right)^{ 1- \beta }}\nonumber\\
        &+\left( {1 - \beta } \right) \sum\limits_{n \in \mathcal{N}} \frac{{{\left(  {P_{n}^{\max }} \right)}^\beta }}{{{\eta^{\text{max}}}}}{\left( {\sum\limits_{k \in \mathcal{K}} { {{{\left| {\boldsymbol{1}_n^{{N}}\boldsymbol{\bar w}_{k}^{\left( j \right)}} \right|}^2}} } } \right)^{ \!\!-\! \beta }}\nonumber\\
        &\cdot \sum\limits_{k \in \mathcal{K}}\left(  { {{{\left| {\boldsymbol{1}_n^{{N}}\boldsymbol{w}_{k}} \right|}^2}} }   -  {{{\left| {\boldsymbol{1}_n^{{N}}\boldsymbol{\bar w}_{k}^{\left( j \right)}} \right|}^2}}    \right). 
\end{align}
 \end{small}Similarly, the approximate RF chain power consumption term $\hat P_{\text{RF}}(\{\boldsymbol{w}_{k}\})$ in \eqref{RF_power_consum_re} is approximated by its upper bound $\tilde{P}_{\text{RF}}^{(j)}(\{\boldsymbol{w}_{k}\})$, i.e.,
 \begin{small}
\begin{align}
        &\hat P_{\text{RF}}(\{\boldsymbol{w}_{k}\}) \leq \tilde{P}_{\text{RF}}^{(j)}(\{\boldsymbol{w}_{k}\})\nonumber\\
        &=\frac{P_{\text{RF}}^{s}}{{\log \left( {1 + \epsilon^{ - 1}} \right)}}\sum\limits_{n \in \mathcal{N}} \left[{\log \left(\!\! 1\!\! +\!  \epsilon^{ - 1}{\sum\limits_{k \in \mathcal{K}} { {{{\left| {\boldsymbol{1}_n^{{N}}\boldsymbol{\bar w}_{k}^{\left( j \right)}} \right|}^2}} } } \!\! \right)} \right.\nonumber\\
        &\left.+{\frac{1}{{ {\sum\limits_{k \in \mathcal{K}} { {{{| {\boldsymbol{1}_n^{{N}}\boldsymbol{\bar w}_{k}^{\left( j \right)}} |}^2}} } }  \!\!\!+\! \epsilon}}\sum\limits_{k \in \mathcal{K}} \left( {{ {{{\left| {\boldsymbol{1}_n^{{N}}\boldsymbol{ w}_{k}} \right|}^2}} } } \!\!\!\!-\!\!{{ {{{| {\boldsymbol{1}_n^{{N}}\boldsymbol{\bar w}_{k}^{\left( j \right)}} |}^2}} } }  \right)} \right]. \label{RFPower_sca_appr} 
\end{align}  
\end{small}

By replacing $P_{\text{PA}}(\{\boldsymbol{w}_{k}\})$ and $\hat P_{\text{RF}}(\{\boldsymbol{w}_{k}\})$ as $\tilde P_{\text{PA}}^{j}(\{\boldsymbol{w}_{k}\})$ and $\tilde P_{\text{RF}} ^{j}(\{\boldsymbol{w}_{k}\})$, respectively, problem (\text{P3}) is transformed into the following problem at iteration $j$ of SCA:
\begin{subequations}
\begin{align}
        &(\text{P4.j}):&{\mathop {\min }\limits_{\left\{ {\boldsymbol{w}_{k}} \right\}} } \quad&  \tilde{P}_{\text{PA}}^{(j)}(\{\boldsymbol{w}_{k}\})  + \tilde{P}_{\text{RF}}^{(j)}(\{\boldsymbol{w}_{k}\}) \nonumber\\
        &&\text{s.t.}\quad&\eqref{P1_cons2}, \eqref{P1_cons3},\eqref{sinr_cons_eq}. \nonumber     
\end{align}
\end{subequations}
Problem (P4.j) is a convex problem that can be solved by convex optimization tools such as CVX \cite{grant2014cvx}. Let $\{\boldsymbol{w}_{k}^{\star(j)}\}$ denote the obtained solution to (P4.j) at the $j$-th iteration, which is then updated as $\{\boldsymbol{w}_{k}^{(j+1)}\}$. Note that as $\tilde P_{\text{PA}}^{j}(\{\boldsymbol{w}_{k}\})$ in \eqref{TransmitPower_sca_appr} and $\tilde P_{\text{RF}} ^{j}(\{\boldsymbol{w}_{k}\})$ in \eqref{RFPower_sca_appr} serve as upper bounds of $P_{\text{PA}}(\{\boldsymbol{w}_{k}\})$ in \eqref{PA_power_consum} and  $\hat P_{\text{RF}}(\{\boldsymbol{w}_{k}\})$ in \eqref{RF_power_consum_re}, respectively, it can be shown that the total power consumption value achieved by $\{\boldsymbol{w}_{k}^{(j+1)}\}$ is always no greater than that by  $\{\boldsymbol{w}_{k}^{(j)}\}$. In other words, the achieved power consumption value of problem (P3) is  monotonically non-increasing after each interation of SCA. Since, the optimal value of problem (P3) is bounded due to the non-negativity of it. Thus, the convergence of the proposed SCA-based algorithm is guaranteed. Let $\{{\boldsymbol{w}_{k}^{\star}}\}$ denote the converged solution to problem (P3).

Finally, we propose a beamforming weight based method to determine the antenna selection on-off selection based on the obtained approximate solution $\{\boldsymbol{w}_{k}^{\star}\}$. First, we define the beamforming weight for each antenna $n$ as $v_{n}=\sum_{k=1}^{K}|{{w}_{k,n}^{\star}}|^2$, where ${w}_{k,n}^{\star}$ is the $n$-th element in ${\boldsymbol{w}_{k}^{\star}}$. The weight factor $v_{n}$, which captures the transmit power at that antenna. It is clear that if $v_{n}$ is small, then antenna $n$ would have a higher priority to be switched off. Next, we sort the beamforming weights of different antennas in an ascending order, and then determine the on-off status of antennas in an iterative manner. In particular, we iteratively switch off the antennas based on the above ordering, and compare their correspondingly achieved power consumption values to obtain the antenna on-off selection that minimized the total power consumption. 

\section{Numerical Results}
This section provides numerical results to validate the effectiveness of our proposed algorithms. In the simulation, we consider the  Rayleigh fading channel model. We set the maximum transmit power as $P_t  = 46$ dBm, the noise power spectrum density as $-174$ dBm/Hz, and the bandwidth as 20 MHz. We also set $P_n^{\text{max}}=1.5$, $P_{\text{RF}}^{s}=0.35$, $P_{c}=20$, and $\eta^{\text{max}}=0.38$.  The SINR constraints at different users are set to be identical, i.e., $\Gamma_k = \Gamma, \forall k \in \mathcal{K}$.

For performance comparison, we consider the following benchmark schemes by considering the fixed PA and/or ignoring the on-off RF chain power consumption, in comparison to our proposed \textbf{joint design with non-linear PA efficiency}. 
\begin{itemize}
\item \textbf{Joint design with fixed PA efficiency}: This corresponds to our proposed joint design by setting $\beta = 0$.
\item \textbf{Beamforming only with non-linear PA efficiency}: This corresponds to the case with $P_{\text{RF}} = N P_{\text{RF}}^s$ being a constant, in which the transmit beamformers $\{\boldsymbol{w}_{k}\}$ can be optimized similarly as in the proposed algorithm to minimize the total power consumption at the BS. 
\item \textbf{Beamforming only with fixed PA efficiency}: This corresponds to the design with $\beta =0$ and $P_{\text{RF}} = N P_{\text{RF}}^s$  being a constant. 
\end{itemize}

\begin{figure}[htbp]
	 \vspace{-15pt}
	\setlength{\abovecaptionskip}{-0pt}
	\setlength{\belowcaptionskip}{-15pt}
	\centering
	\includegraphics[width= 0.42\textwidth]{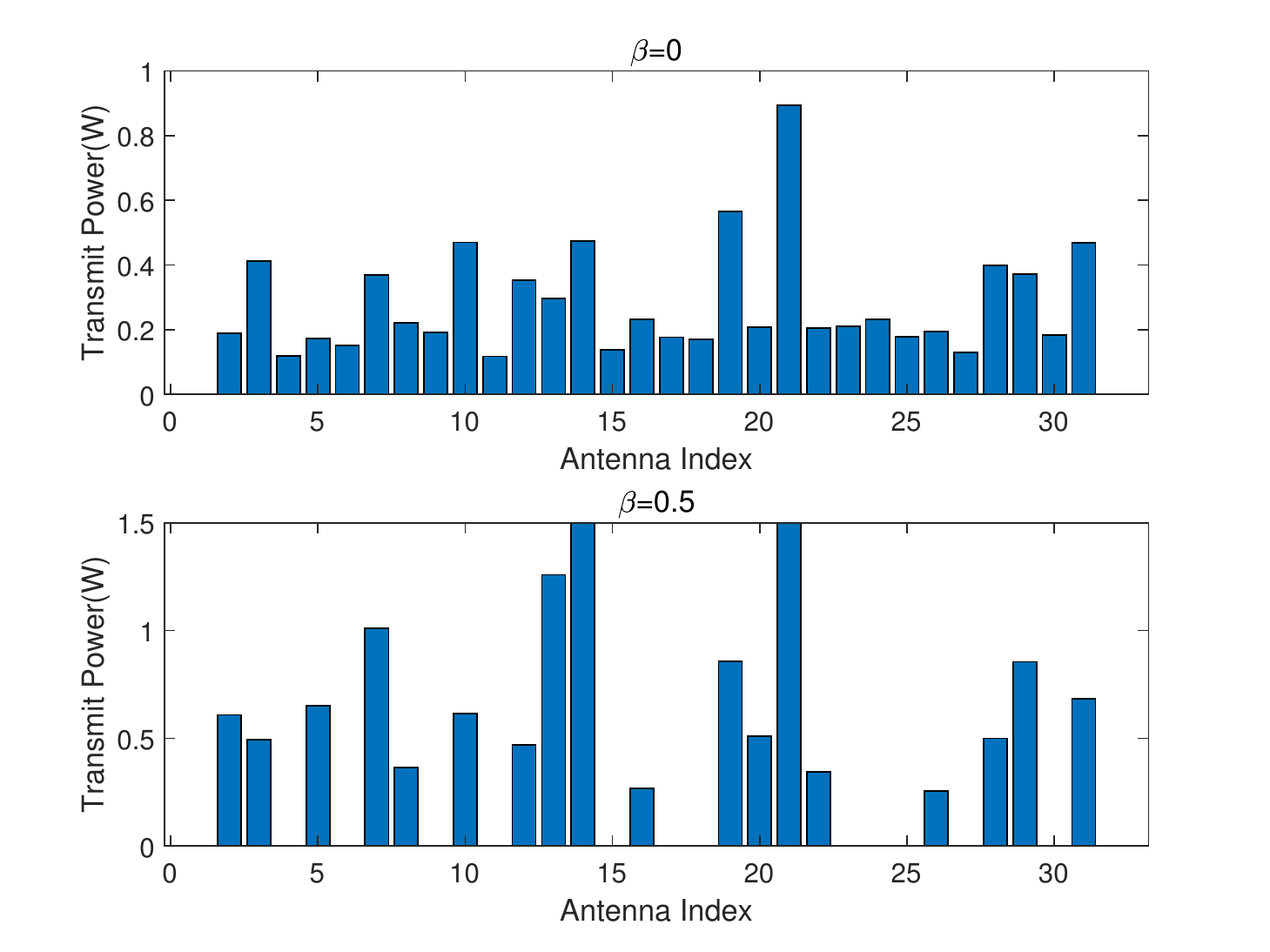}
	\DeclareGraphicsExtensions.
	\caption{The transmit power allocation at different transmit antennas, where $M=32$, $K=8$, and $\Gamma = 12$ dB.}
	\label{PowVsAntSINR12}
\end{figure}
Fig.~\ref{PowVsAntSINR12} compares the optimized transmit power allocation at different antennas in the practical case with nonlinear PA efficiency (i.e., $\beta =0.5$) versus that with fixed PA efficiency (i.e., $\beta =0$). It is observed that in the conventional design with fixed PA efficiency, almost all antennas are switched on with properly allocated transmit power. By contrast, in our proposed design with non-linear PA efficiency, it is observed that some antennas are switched off with zero transmit power, while other antennas are allocated with more power to meet the users' SINR constraints. This is due to the fact that the non-linear PA efficiency becomes higher as the output signal power at each antenna, and as a result, the BS tends to switch off more antennas to save the RF-chain power consumption and increase the transmit power on the remaining antennas to exploit higher PA efficiency.

\begin{figure}[htbp]
	 \vspace{-10pt}
	\setlength{\abovecaptionskip}{-0pt}
	\setlength{\belowcaptionskip}{-10pt}
	\centering
	\includegraphics[width= 0.42\textwidth]{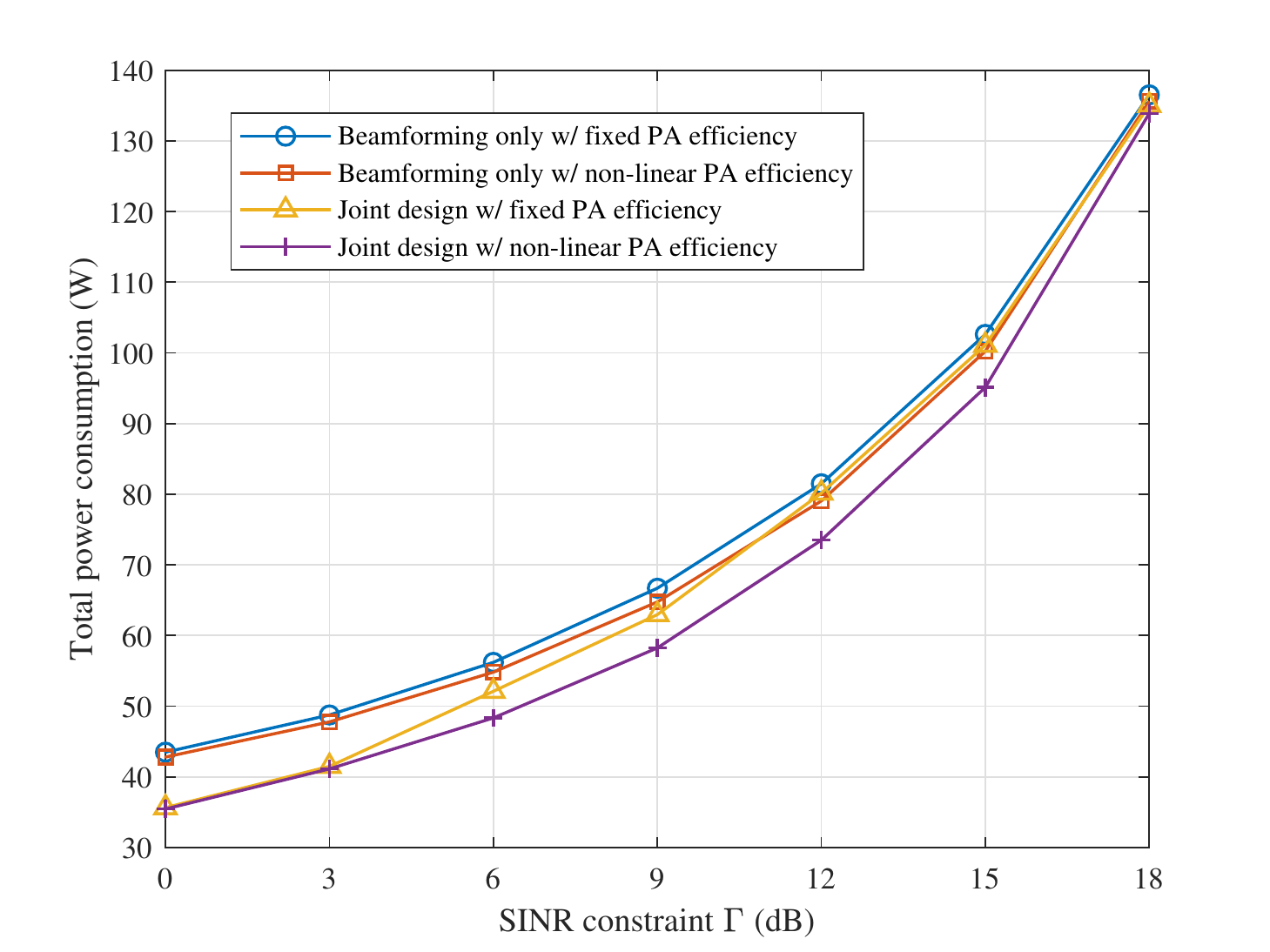}
	\DeclareGraphicsExtensions.
	\caption{The total power consumption versus SINR constraint $\Gamma$ with $M=32$ and $K=8$.}
	\label{PowerVsSINR}
\end{figure}
Fig.~\ref{PowerVsSINR} shows the total power consumption at the BS versus the SINR constraint $\Gamma$ at each user, where $M=32$ and $K=8$. It is observed that in the whole SINR regime, our proposed joint design with non-linear PA efficiency achieves the lowest power consumption among all schemes. In the low SINR regime, the joint design with fixed PA efficiency is observed to perform close to that with non-linear PA efficiency, and outperforms the beamforming only design with non-linear or fixed PA efficiency. This is because in this case, activating a small number of antennas is sufficient to meet the SINR constraints, and thus antenna on-off selection is critical for power saving. Furthermore, in the high SINR regime, it is observed that the power consumption values of the four schemes become similar.  This is because in this case, the BS has to activate all antennas with full transmit power to meet the stringent SINR requirements at users.

\begin{figure}[htbp]
	\vspace{-10pt}
	\setlength{\abovecaptionskip}{-0pt}
	\setlength{\belowcaptionskip}{-10pt}
	\centering
	\includegraphics[width= 0.42\textwidth]{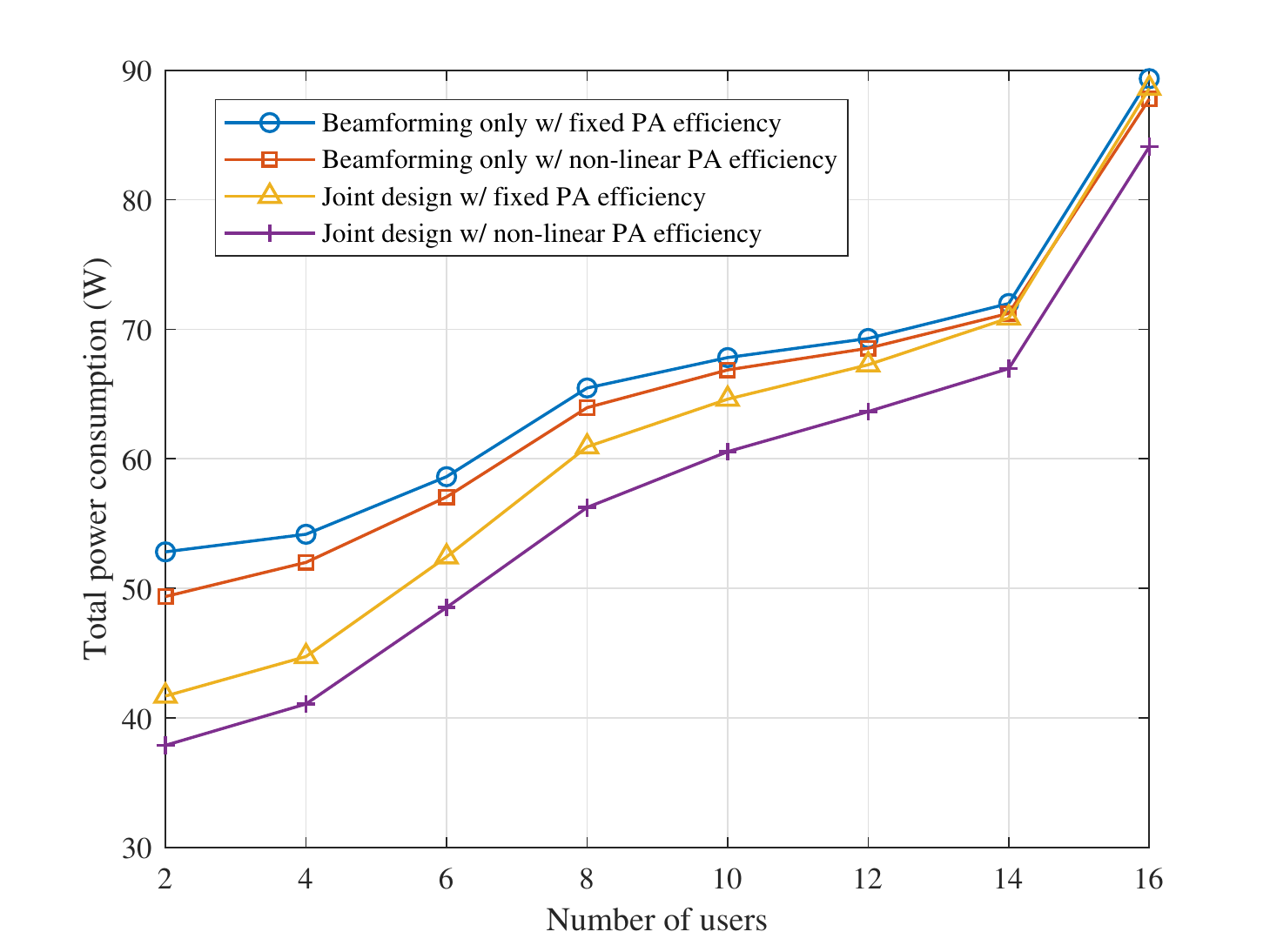}
	\DeclareGraphicsExtensions.
	\caption{The total power consumption versus the number of users $K$ with $M=32$ and $\Gamma = 3$ dB.}
	\label{PowerVsUserNum_SINR3}
\end{figure}
Fig.~\ref{PowerVsUserNum_SINR3} shows the total power consumption at the BS versus the number of users $K$, where $M = 32$ and $\Gamma = 3$ dB. When the number of users $K$ is small, it is observed that the joint design with nonlinear PA efficiency and that with fixed PA efficiency outperform the beamforming only designs. This is due to the fact that the BS can switch off more antennas and RF chains for minimizing the total power consumption. When K becomes large, it is observed that the joint design with non-linear PA efficiency outperforms the other three benchmark schemes. This is because that in the proposed joint design, the BS tends to use less antenna and higher transmit power at each antenna to improve the PA efficiency, thus leading to more significant power saving.

\begin{figure}[htbp]
	 \vspace{-10pt}
	\setlength{\abovecaptionskip}{-0pt}
	\setlength{\belowcaptionskip}{-10pt}
	\centering
	\includegraphics[width= 0.42\textwidth]{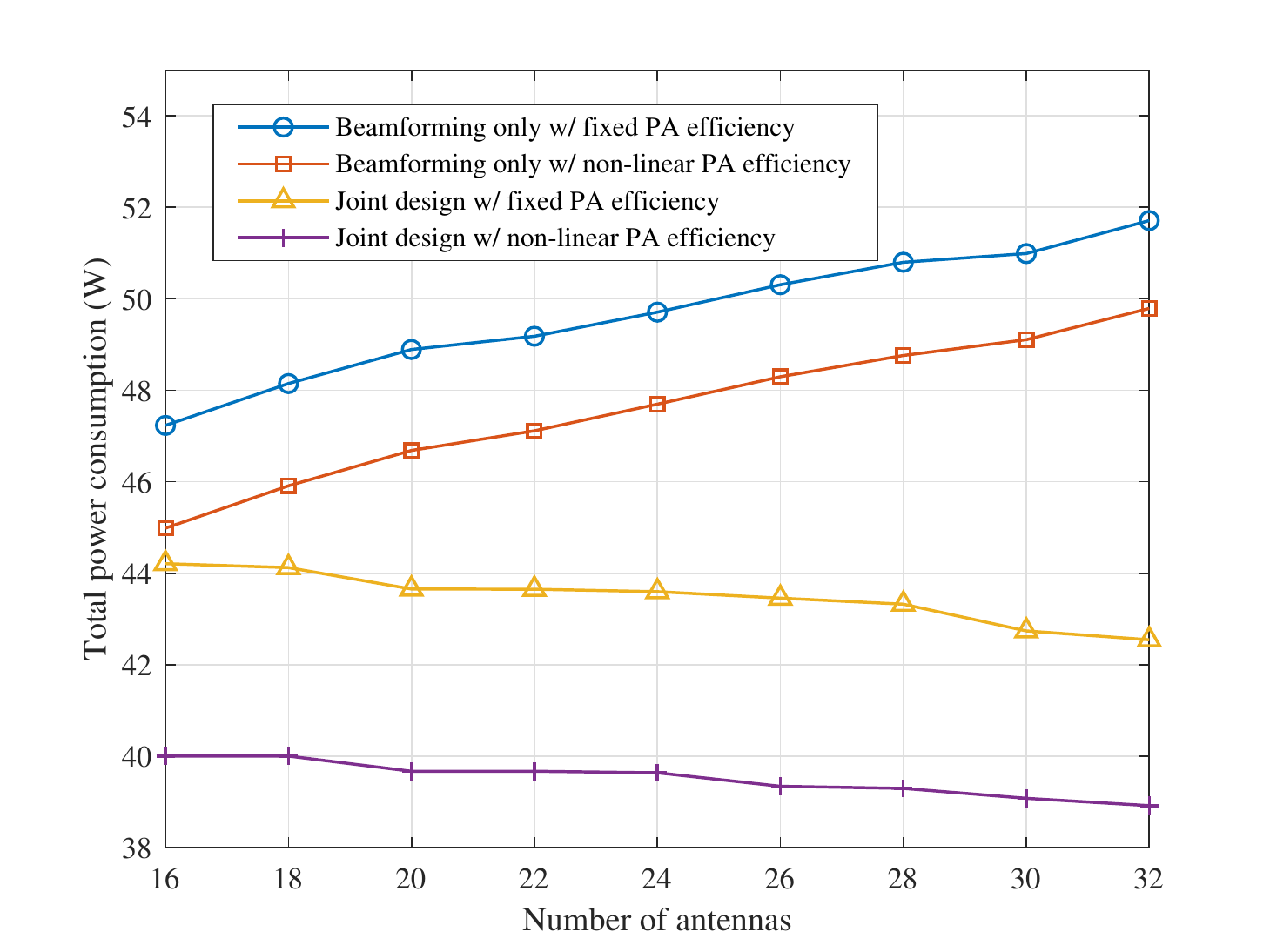}
	\DeclareGraphicsExtensions.
	\caption{The total power consumption versus the number of antennas $M$ with $K=4$ and $\Gamma = 3$ dB.}
	\label{PowerVsAntNum_SINR3}
\end{figure}
Fig.~\ref{PowerVsAntNum_SINR3} shows the total power consumption versus the number of antennas at BS $M$, where $K=4$ and $\Gamma=3$. It is observed that for the beamforming only designs, the total power consumption at the BS increases as the number of antennas M becomes large. In this case, the increasing power consumption mainly comes from the increased number of active RF chains. By contrast, it is observed that for the joint design, the total power consumption at the BS is monotonically non-increasing as $M$ increases. This is because that in the joint design, the BS can properly select the active antennas to increase the communication performance without increasing the RF-chain power consumption.
\section{Conclusion}
This letter studied the energy-efficient beamforming and antenna selection in a multi-antenna multiuser communication system by considering both the non-linear PA efficiency and the on-off RF-chain power consumption. In order to minimize the total power consumption at the BS while ensuring the SINR constraints at users, we proposed an efficient transmit beamforming design algorithm by using the SCA and an antenna selection algorithm based on beamforming weights. Numerical results showed that the proposed algorithm significantly reduces the total power consumption at the BS as compared to the conventional designs by considering the fixed PA efficiency and/or ignoring the on-off RF-chain power consumption. In particular, it was shown that the BS tends to activate fewer antennas with higher transmit power rather than more antennas with lower power transmission to exploit the non-linear PA efficiency and on-off RF chains for power saving.

\bibliographystyle{IEEEtran}
\bibliography{IEEEabrv,myref}

\end{document}